\documentstyle[aps,prl,twocolumn,epsf]{revtex}
\begin{document}

\title{Manifestation of spin--charge separation in
the dynamic dielectric response of one--dimensional Sr$_2$CuO$_3$
}

\author{R. Neudert, M. Knupfer, M. S. Golden, and J. Fink}
\address{Institut f\"ur Festk\"orper- und Werkstofforschung Dresden, P.O.Box
270016,
D-01171 Dresden, Germany
}

\author{W. Stephan}
\address{Max-Planck-Institut f\"ur Physik komplexer Systeme, N\"othnitzer
Str. 38,
D-01187 Dresden, Germany
}

\author{K. Penc}
\address{Service de Physique Th\'eorique, CEA Saclay, 91191 Gif-sur-Yvette
Cedex,
France
}

\author{N. Motoyama, H. Eisaki, and S. Uchida}
\address{Department of Superconductivity, The University of Tokyo,
Bunkyo-ku,
Tokyo 113, Japan
}

\date{April 2, 1998}

\maketitle

\begin{abstract}
We have determined the dynamical dielectric response of a one--dimensional,
correlated insulator by carrying out electron energy--loss spectroscopy on
Sr$_2$CuO$_3$ single crystals.
The observed momentum and energy dependence of the low--energy features,
which correspond to collective transitions across the gap,
are well described by an extended one--band Hubbard model with moderate
nearest neighbor Coulomb interaction strength.
An exciton--like peak appears with increasing momentum transfer.
These observations provide experimental evidence for spin--charge
separation in the relevant excitations of this compound, as theoretically
expected for the one--dimensional Hubbard model.
\end{abstract}


Low dimensional, correlated systems have attracted much attention in recent
years not only because of the discovery of high-temperature
superconductivity
for the two-dimensional (2D) case of copper-oxide planes, but in general due
to a large variety of unconventional magnetic and electronic properties
directly connected with the reduced dimensionality and/or electronic
correlations.
In this context one--dimensional (1D) systems are of special interest,
essentially stimulated by theoretical predictions like that of spin--charge
separation for the Hubbard model
\cite{spinChargeseparation,spinChargefactorization}.
Quasi--1D materials based on cuprate compounds, of which Sr$_2$CuO$_3$ is
the best example, have become new candidates for ideal model
systems which allow the study of basic physical concepts in one--dimension
and
represent a touchstone for the theoretical models of high-T$_{\rm
c}$-cuprates.
Magnetic susceptibility measurements have shown that Sr$_2$CuO$_3$ can be
regarded as an almost ideal realization of the 1D spin-$\frac{1}{2}$
antiferromagnetic (AF) Heisenberg model\cite{Ami95,Moto96},
which describes the magnetic excitations of a Mott--Hubbard insulator.

Information on the electronic structure and the dynamics of the charge
carriers is highly desirable, especially against the background of
spin--charge separation expected in 1D. Up to now angle resolved
photoemission
spectroscopy (ARPES) has been performed on the related compound SrCuO$_2$
\cite{Kim96}, which is composed of two neighboring, but at room temperature
magnetically decoupled, copper-oxide chains. The ARPES data have been
interpreted in terms of holon and spinon bands, with bandwidths related to
the hopping term $t$ and the exchange constant $J$, respectively.
Beside the one--particle spectral function obtained by photoemission, the
dielectric function is the most basic and important quantitiy reflecting the
electronic structure of a solid. The dielectric response is accessible using
optical spectroscopy for the special case of zero momentum transfer.
Electron energy--loss spectroscopy (EELS) on the other hand,
offers the possibility to study the momentum dependence
of the electronic excitations, i.e. the dynamical dielectric response.
EELS measurements for the 2D system Sr$_2$CuO$_2$Cl$_2$ \cite{Wang}
have been interpreted in terms of a small exciton model,
where it was assumed that the singlet excitons may propagate
freely in the AF spin background, whereas single particle propagation
is suppressed. In contrast, in our 1D case spin--charge separation naturally
leads to no frustration of the kinetic energy of carriers excited across the
gap, and therefore both continuum states as well as excitonic bound states
play a role.

In this Letter we present the first investigations of the dynamical
dielectric response of Sr$_2$CuO$_3$.
We have carried out EELS measurements in transmission on single crystalline
samples, which provides us with the energy and momentum dependent loss
function Im$(-1/\epsilon(\vec{q},\omega))$.
While for small momentum transfer we see a broad continuum of interband
plasmons above the gap, on the way to the zone boundary
a sharp peak develops.  We show that the data can be understood within
an extended effective one--band Hubbard model, and that
both the spin--charge separation which occurs in 1D as well
as excitonic effects are essential.

Single crystals of Sr$_2$CuO$_3$ were grown using a travelling solvent-zone
technique \cite{Moto96}. For our EELS studies in transmission, films of
about 1000\,{\AA} thickness were cut from the crystals using an
ultramicrotome
with a diamond knife. The measurements were performed at room temperature
using a specially designed high resolution spectrometer \cite{Fink89}
with a primary beam energy of 170\,keV. The energy and momentum resolution
were set to be 115\,meV and 0.05\,\AA$^{-1}$ for $q\le$0.5\,\AA$^{-1}$,
and 160\,meV and 0.06\,\AA$^{-1}$ for $q\ge$0.5\,\AA$^{-1}$
due to the necessity of increasing the electron beam intensity
for large momentum transfer $q$. The high quality and the orientation
of the single crystalline films were carefully checked by {\em in situ}
electron diffraction. For recording the loss function the momentum transfer
was aligned along the chain direction ([010]). It is important to note that
EELS in transmission is a not surface sensitive technique, in contrast to
many other electron spectroscopies. The dielectric function,
$\epsilon(\vec{q},\omega)=\epsilon_1+i\epsilon_2$, and thus the
optical conductivity $\sigma=\omega\epsilon_2$ was obtained from the
loss function by a Kramers--Kronig analysis.

\begin{figure}
 \epsfxsize=8.5 truecm
 \centerline{\epsffile{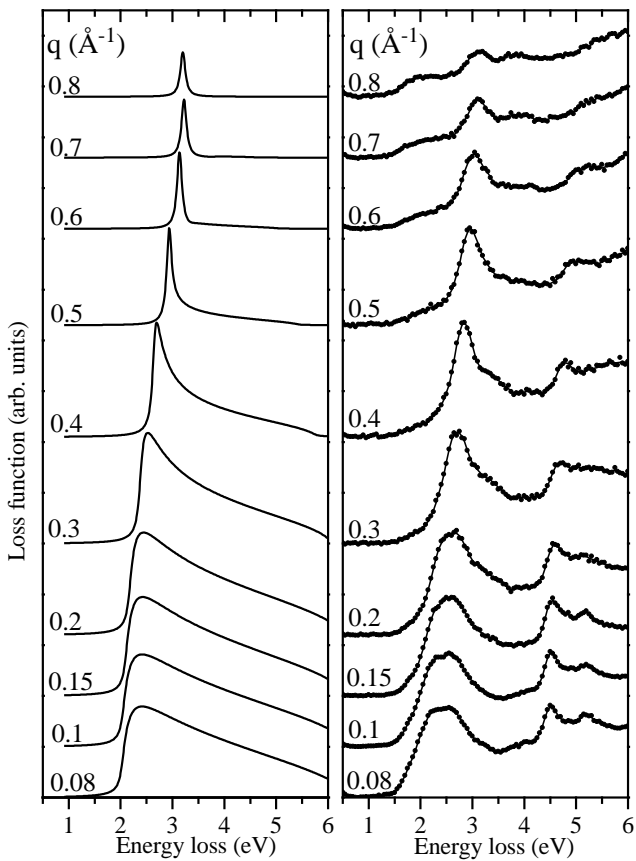}}
\caption{Loss function of Sr$_2$CuO$_3$ (right panel) measured with
 the momentum transfer $\vec q$ parallel to the chain direction.
 The left panel shows the calculated loss function ${\cal N}(q,\omega)$
 plotted with an energy resolution of 0.115\,eV, and scaled to the
experimental
 peak heights. For the parameters used in the calculation, see the text.
}
\end{figure}

In the right panel of Fig. 1 we show the loss function of Sr$_2$CuO$_3$
for different momentum transfers $q$ parallel to the chain direction.
The spectra are normalized to equal count rates in the energy range of
9.4-10.4\,eV (not shown), where they are practically momentum independent.
Due to contributions of the elastic line and surface losses it is not
possible
to measure at zero momentum transfer but very close to the optical limit
($q=$0.08\,\AA$^{-1}$). With $q=$0.8\,\AA$^{-1}$ being the zone boundary,
our measurements cover the complete Brillouin zone in the [010] direction.
Peaks in the loss function in the low energy range discussed here
arise from collective excitations (plasmons) related to interband
transitions.
In the spectrum for $q=$0.08\,\AA$^{-1}$, the first possible interband
transitions across the gap form a broad continuum like absorption feature
in the loss function around 2.4\,eV.
The spectral onset is found to be at 1.6\,eV. Two further comparatively
sharp
maxima are observed at 4.5\,eV and 5.2\,eV, which are probably already
connected with excitations not only within the CuO$_3$ chain, but involving
Sr orbitals. As a function of momentum transfer, the peak at 4.5\,eV shifts
towards higher energy, accompanied by a drop in intensity and an increasing
spectral width. This mainly reflects the decreasing lifetime
of the excited state and represents the usual behavior of an interband
plasmon in EELS.

In the following we will focus on the loss function in the range of 1.6\,eV
to $\sim$4\,eV since here the spectral features are expected to be
exclusively due to transitions within the CuO$_3$ chain and model
calculations
are believed to be of relevance. When comparing the momentum dependence of
the
loss function for this energy range with the higher lying features described
above a completely different behavior is observed.
The broad absorption continuum present near the zone center
($q=0.08$\AA$^{-1}$) narrows with increasing momentum transfer, evolving
into
a single rather sharp peak centered at 2.8\,eV for $q=0.4$\AA$^{-1}$ with a
width of only $\sim 0.5$\,eV. At the same time, the integrated intensity
stays
almost constant indicating a transfer of spectral weight from the continuum
to
the sharp peak. The remainder of the continuum is visible as a
shoulder at around 3.3\,eV. Going into the second half of the Brillouin zone
the peak gets slightly broader possibly due to an enhanced total background
arising from multiple scattering for large $q$. At the zone boundary
($q=0.8$\AA$^{-1}$) the excitation is centered at 3.2\,eV. Additionally,
close
to the zone boundary a second peak appears at $\sim$3.8\,eV. The spectral
weight at 2\,eV  which starts to be visible for $q\ge0.5$\AA$^{-1}$ has its
origin in double scattering processes (inelastic scattering plus phonon or
impurity scattering with $q$ equal to that set by the spectrometer), leading
to the appearance of the spectrum for
$q=0$ in the curves at high momentum transfer \cite{Fink89}.

The most unusual feature of the loss function described above is
the anomalous narrowing of the low--energy peak with increasing momentum
transfer.  Although it is widely accepted that cuprates like Sr$_2$CuO$_3$
are
charge transfer insulators,
an effective one--band Hubbard model has been quite successful
in describing many low--energy experimental features of related compounds
\cite{Kim96,1band}.  One is therefore led to ask whether or not this
also holds here.
Recently, two of us \cite{Step96} studied the imaginary part of the
zero temperature density response function
\begin{equation}
  {\cal N}_0(q,\omega) = \sum_f
     \biglb| \langle f | n_{q} | 0 \rangle \bigrb|^2
     \left(
	\frac{1}{\omega \!-\! \varepsilon_{f0} \!+\! i\delta}
       -\frac{1}{\omega \!+\! \varepsilon_{f0} \!+\! i\delta}
     \right)
\label{nkw}
\end{equation}
of the extended one--band Hubbard model,
\begin{eqnarray}
 {\cal H} &=& -t \sum_{j,\sigma}
  \left(
    c^{\dagger}_{j+1,\sigma} c^{\phantom{\dagger}}_{j,\sigma}
    + {\rm h.c.}
  \right)
  + U \sum_{j} n_{j,\uparrow} n_{j,\downarrow} \nonumber\\
  &&+ V \sum_{j} n_{j} n_{j+1}
\label{xhubbard}
\end{eqnarray}
where $n_i=n_{i\uparrow}+n_{i\downarrow}$, with Fourier transform $n_q$,
$\varepsilon_{f0}=E_f-E_0$ and $| 0 \rangle$ is the ground state.
The local Coulomb interaction $U$ of this model would correspond to the
charge transfer energy of a multi--band charge transfer model.
The loss function is proportional to the imaginary part of the
response function shown in Eq. (\ref{nkw}),
if the long--range Coulomb interaction $V_{\rm 3D}(q)=4\pi e^2/q^2$
were also included in the model.
We may, however, include this at an RPA-level of approximation by using
the response function calculated for the short--range interaction model
(\ref{xhubbard}) as the
``Lindhard function'' within the RPA \cite{RPA}:
\begin{equation}
  {\cal N}(q,\omega) =
     \frac{{\cal N}_0(q,\omega)}
          {1-V_{\rm 3D}(q){\cal N}_0(q,\omega)}.
\label{NRPA}
\end{equation}
Within this approach we find that the screening effects can be quite well
described as a renormalization of the short--range interaction, so that a
qualitatively correct picture is provided already with the short range
model.

As a first step we therefore want to discuss our experimental data in a
qualitative way making use of the result for the density response function
of the model (\ref{xhubbard}).
It was shown in Ref. \onlinecite{Step96} that in the strong--coupling
limit $U/t \gg 1$,  the response function ${\cal N}_0(k,\omega)$
may be calculated within an effective $t-J$--like model, where exactly
one ``hole'' and one ``double occupancy'' are explicitly included in the
states excited across the Mott--Hubbard gap.
It was further shown that upon making use of the wave function factorization
\cite{spinChargefactorization} which holds in this limit, to very good
approximation the spin degrees of freedom decouple from the problem, which
is
due to the off-diagonal long range order of singlet pairs in the quantum
spin
chain\cite{talstra95}. We are then left with an effective (spinless)
particle--hole model, with a nearest neighbor attraction $V$ between
the double occupancy (doublon) and holon, with opposite signs of hopping
matrix element for the two carriers, and with the band centers separated
by $U$.  As sketched in Fig. 2, in the small $V$ limit one then expects
an optical gap of $U-4t$, followed by a continuum of interband transitions
up to an energy of $U+4t$. Going to higher momentum transfers the range of
possible interband transitions decreases, leading to an excitation energy
$U$
at the zone boundary.
The inclusion of  finite nearest-neighbor Coulomb repulsion $V$ leads to the
possibility of the formation of an excitonic state. For  $V<2t$ the exciton
lies within the continuum in the optical limit ($q \rightarrow 0$), and is
therefore not a well defined excitation there, but will appear at the zone
boundary at energy $U-V$, accounting for almost all of the spectral weight
(Fig. 2 (b)) \cite{Step96}. The narrowing of the low-energy feature in the
EELS data with increasing $q$ can be explained within this scheme.
Once again we stress that the effective band structure sketched in Fig. 2
(a)
represents the holon--doublon dispersion relations, and not a band structure
in the conventional sense. If there were significant coupling to spin
excitations then the narrowing of the continuum would be
counteracted by the possiblity of momentum transfer to spinons.

\begin{figure}
 \epsfxsize=8.5 truecm
 \centerline{\epsffile{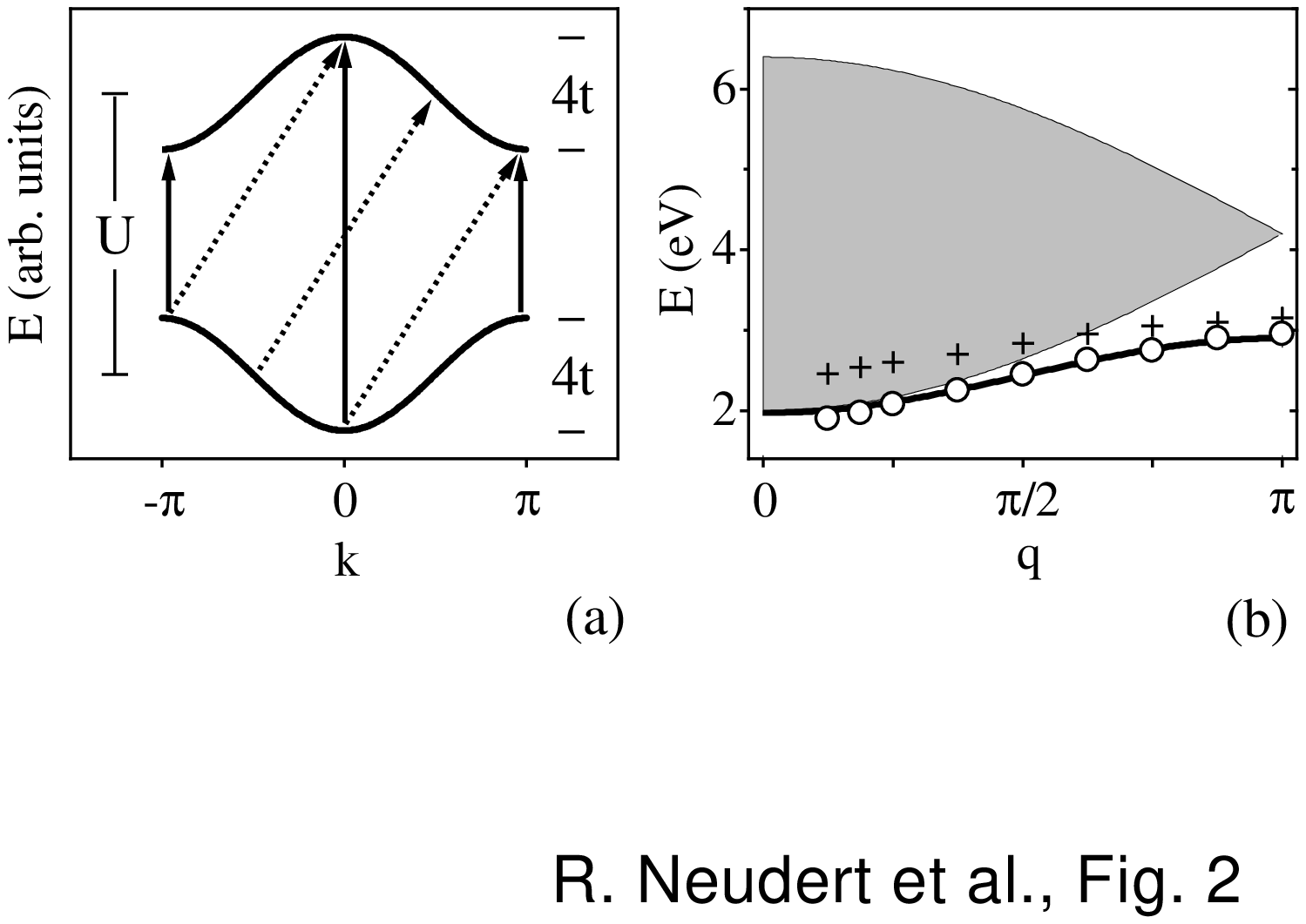}}
\caption{(a) Schematic representation of the effective band structure of a
 1D Mott-Hubbard insulator after decoupling of spin degrees of freedom.
 Also depicted are examples of possible interband transitions for momentum
 transfer $q=0$ (solid arrows) and $q=\pi$ (dashed arrows). The value of
 $q=\pi$ corresponds to $q=0.8$\,\AA$^{-1}$  in the experiment.
 (b) Possible energy range of interband transitions as a function of
momentum
 transfer q (shaded area) and exciton dispersion (heavy line) as obtained
from
 the peak positions of the calculated optical conductivity using the same
 parameter set as in Fig. 1. $\circ$: Peak positions of the optical
conductivity
 as derived from the measured loss function by Kramers-Kronig analysis.
 $+$: dispersion of the low--energy peak of the loss function.
}
\end{figure}

We would now like to make this discussion more quantitative.
To account for the collective excitations excited in an EELS experiment,
the effects of long range Coulomb interaction and interchain coupling
must be included.  As mentioned above, this may be done
within RPA approximation by solving for the density response
${\cal N}_0(q,\omega)$ of the
model (\ref{xhubbard}), and then using the RPA form \cite{RPA} to connect
this screened response to the loss function which is proportional to
Im${\cal N}(q,\omega)$, Eq. (\ref{NRPA}).

In Fig. 1 (left panel) we present the loss function calculated using the
solution for ${\cal N}_0(q,\omega)$ given in Ref. \onlinecite{Step96}
and Eq. (\ref{NRPA}). The curves shown here are obtained with the parameter
set $t=0.55$\,eV, $U=4.2$\,eV, $V=1.3$\,eV, and represent the best
description of the experimental data.  Although a quantitative
error analysis is difficult, already a 10 percent deviation from
these values leads to noticeably worse agreement with the measurements.
Note that the momentum dependence of the lineshape provides a strong
constraint on the parameters in addition to that given by the dispersion.
Of course, the exact lineshape observed at low $q$ is not reproduced by our
theory, but considering the simplifications inherent in our model
the agreement with the measured loss function (right panel) is reasonable.

Our value for $t$ is consistent with $t=0.6$\,eV obtained from the holon
dispersion observed in SrCuO$_2$ in Ref. \onlinecite{Kim96}, as well as
$t=0.55$\,eV extracted from band structure calculations \cite{Rosn97}.
Once again, we must emphasize that $V=1.3$\,eV here represents the
{\em unscreened} value, and that screening effects are treated in the RPA.
For a Hamiltonian of the form (\ref{xhubbard}) without long--range Coulomb
interactions, a smaller screened value of $V$ is appropriate.
In the present case $V \approx 0.8$\,eV used with Eqs. (\ref{nkw}) and
(\ref{xhubbard}) leads to a similar correspondence to our EELS data, and
should be used for model studies without long range interactions.
In this context we note that a possible screening of U would be a higher
order
effect and thus would be unlikely to play a significant role.

To obtain more direct information about the interband transitions
responsible
for the collective excitations in the loss function we have also derived the
optical conductivity $\sigma$ from our measured data by a Kramers--Kronig
analysis with $\epsilon_1(\vec q, \omega=0)=8$.
For small momentum transfer our result is consistent with optical
measurements
\cite{Mait97} of the conductivity, where no indications of excitonic states
are observed, despite the much better energy resolution.
This corroborates our interpretation of the peak at small momentum transfer
being a continuum, and not for example as resulting from several overlapping
peaks which are not resolved clearly.  Regarding the dispersion, the peak
position in the optical conductivity as obtained from our EELS data is in
excellent agreement with the corresponding peaks  of $\sigma$ directly
calculated within the above model and parameter set (heavy line
in  Fig. 2 (b)).
In addition, the peak positions of the loss function (denoted by crosses)
clearly show the expected plasmonic shift relative to those in $\sigma$.

While this work demonstrates the applicability of the effective one--band
model for the low--energy electronic properties of the Cu-O chains, our
treatment here is clearly semi--phenomenological in the sense that
the model itself is microscopically non--trivial to justify.
When one considers more complicated multi--band models which
more obviously reflect the local chemistry of Sr$_2$CuO$_3$,
excitations appear at higher energies which are not included within our
present approach. In general some of the degrees of freedom included in
multi--band models may lead to relevant additional excitations
overlapping in energy range with those described by the simple
one--band model.
Calculations in the framework of a two-band model show additional
``excitonic'' features for large momentum transfers beyond the one obtained
in the one--band model, but with very similar low--energy and small momentum
transfer behavior to that discussed here \cite{Penc98}. The peak at 3.8\,eV
observed close to the zone boundary in EELS can probably be described within
a two- or three-band model, although the results of Ref. \onlinecite{Penc98}
are quantitatively accurate only for parameters which are not realistic
for Sr$_2$CuO$_3$, so this point remains open.

In conclusion, we have carried out EELS measurements of the one--dimensional
correlated insulator Sr$_2$CuO$_3$.  The momentum and energy dependence of
the dielectric response at low energies can be well described within an
extended one--band Hubbard model. The unusual narrowing of the lowest energy
feature with increasing momentum transfer arises due to a combination of two
effects:
a) the band structure of the excited carriers in the effective
model quite naturally leads to a kinematic narrowing of the
interband continuum with increasing momentum;
b) the presence of a moderately large Coulomb attraction between the
excited holon--doublon pair leads to an excitonic bound state, which
sharpens and lies below the continuum only for momentum transfers away from
the Brillouin zone center.
Due to the spin--charge factorization which holds for our 1D model, we are
able to discuss in points a) and b) a model with no coupling of carriers to
the spin background. Therefore the measurements presented here are also a
manifestation of the charge--spin separation theoretically expected for the
1D Hubbard model.

We would like to acknowledge fruitful discussions with
S.-L. Drechsler and H. Shiba.  Financial support was provided by the
German {\it Bundesministerium f\"ur Bildung, Forschung und Technologie}
(BMBF)
under Contract No. 13N6599/9, the {\it Max Planck Gesellschaft}, the
{\it Ministry of Education, Science and Culture, Japan} (COE Grant), and
the {\it New Energy and Institute Technology Developement Organization}
(NEDO).

\end{document}